\documentclass[prb,onecolumn,12pt]{revtex4-1}
\usepackage{epsfig}
\usepackage{amssymb}
\usepackage{threeparttable}
\usepackage{graphicx}
\usepackage{natbib}
\usepackage{longtable}
\usepackage{txfonts}
\usepackage{color}

\begin{document}
\title{Theoretical investigation of novel electronic, optical, mechanical and thermal properties of metallic hydrogen at 495 GPa}
\author{Bo Peng$^1$, Ke Xu$^1$, Hao Zhang$^{1*}$, Hezhu Shao$^{2}$, Gang Ni$^1$, Jing Li$^1$, Liangcai Wu$^3$, Hongliang Lu$^4$, Qingyuan Jin$^1$, and Heyuan Zhu$^1$}
\affiliation{$^1$Key Laboratory of Micro and Nano Photonic Structures (MOE), Department of Optical Science and Engineering, Fudan University, Shanghai 200433, China\\
$^2$Ningbo Institute of Materials Technology and Engineering, Chinese Academy of Sciences, Ningbo 315201, China\\
$^3$College of Science, Donghua University, Shanghai 201620, China\\
$^4$State Key Laboratory of ASIC and System, Institute of Advanced Nanodevices,School of Microelectronics, Fudan University, Shanghai 200433, China\\
}

\begin{abstract}
Atomic metallic hydrogen has been produced in the laboratory at high pressure and low temperature, prompting further investigations of its different properties. However, purely experimental approaches are infeasible because of the extreme requirements in producing and preserving the metastable phase. Here we perform a systematic investigation of the electronic, optical, mechanical and thermal properties of $I4_1/amd$ hydrogen at 495 GPa using first-principles calculations. We calculate the electronic structure and dielectric function to verify the metallic behaviour of $I4_1/amd$ hydrogen. The calculated total plasma frequency from both intraband and interband transitions, 33.40 eV, agrees well with the experimental result. The mechanical properties including elastic stability and sound velocity are also investigated. The mechanical stability of $I4_1/amd$ hydrogen is limited by shear modulus other than bulk modulus, and the high Young's modulus indicates that $I4_1/amd$ hydrogen is a stiff material. After investigating the lattice vibrational properties, we study the thermodynamical properties and lattice anharmonicity to understand thermal behaviours in metallic hydrogen. Finally, the lattice thermal conductivity of $I4_1/amd$ hydrogen is calculated to be 194.72 W/mK and 172.96 W/mK along the $x$ and $z$ directions, respectively. Using metallic hydrogen as an example, we demonstrate that first-principles calculations can be a game-changing solution to understand a variety of material properties in extreme conditions.
\end{abstract}

\maketitle

\section{Introduction}

In 1935, Wigner and Huntington predicted that, hydrogen molecules would become a metallic solid similar to the alkalis at high pressure \cite{Wigner1935}. Metallic hydrogen is expected to be a wonder material: it may be a room-temperature superconductor and have significant applications in energy and rocketry \cite{Ashcroft1968}. However, producing metallic hydrogen has been a great challenge. Although a lot of theoretical investigations have been performed on metallic hydrogen \cite{Cudazzo2010a,McMahon2011,McMahon2011a,McMahon2012,Azadi2014,McMinis2015,Borinaga2016,Borinaga2018}, not until recently has atomic metallic hydrogen been synthesized in the laboratory at high pressure and low temperature \cite{Dias2017}. At a pressure of 495 GPa, hydrogen becomes metallic with a high reflectivity, and the metallic phase is metastable when the pressure is released. This prompts a systematic study of its electronic, optical, mechanical and thermal properties for further applications.

However, the challenge is to produce and preserve such metastable phase for experimental investigations. Several months after the synthesis, the world's only metallic hydrogen sample has disappeared. Even if the reproducibility of the sample is improved, characterizing metallic hydrogen under such pressure is extremely difficult due to the limitations of conventional techniques. For instance, for hydrogen samples in a diamond anvil cell at extremely high pressure, neutron scattering and X-ray diffraction experiments are unavailable. Thus it is imperative to develop alternative techniques for exhaustive characterization. Theoretical estimation, as a reference to the experiments, is indeed another option. Recently, computational material techniques have been developed to perform an \textit{ab initio} study on materials in extreme conditions \cite{Burakovsky2010,Gao2008,Alfe2002,Zhu2017,Moustafa2017,Ma2009a}. In addition, state-of-the-art techniques can be used to calculate accurately the ground- and excited-state \cite{Shishkin2006,Shishkin2007,Fuchs2007}, mechanical \cite{LePage2002,Wu2005}, and phononic properties \cite{ShengBTE,Li2012a,Li2012,Li2013a}, enabling the feasibility of theoretical predictions in a variety of different properties of metallic hydrogen.

Here, we perform an exhaustive study of metallic hydrogen at 495 GPa. Our fully first-principles calculations shed light into its electronic, optical, mechanical and thermal properties. First we focus on ground- and excited-state properties such as electronic band structure, dielectric function and plasma frequency. Then, mechanical properties including elastic anisotropy and sound velocity are investigated. After that, we discuss the lattice vibrational properties as well as lattice anharmonicity. Finally we study the phonon transport properties of metallic hydrogen. Our results provide clear means of characterizing metallic hydrogen via these distinct features, strongly calling for experimental verifications.

\section{Ground- and excited-state properties}

\subsection{Crystal structure}

\begin{figure}[h]
\centering
\includegraphics[width=0.45\linewidth]{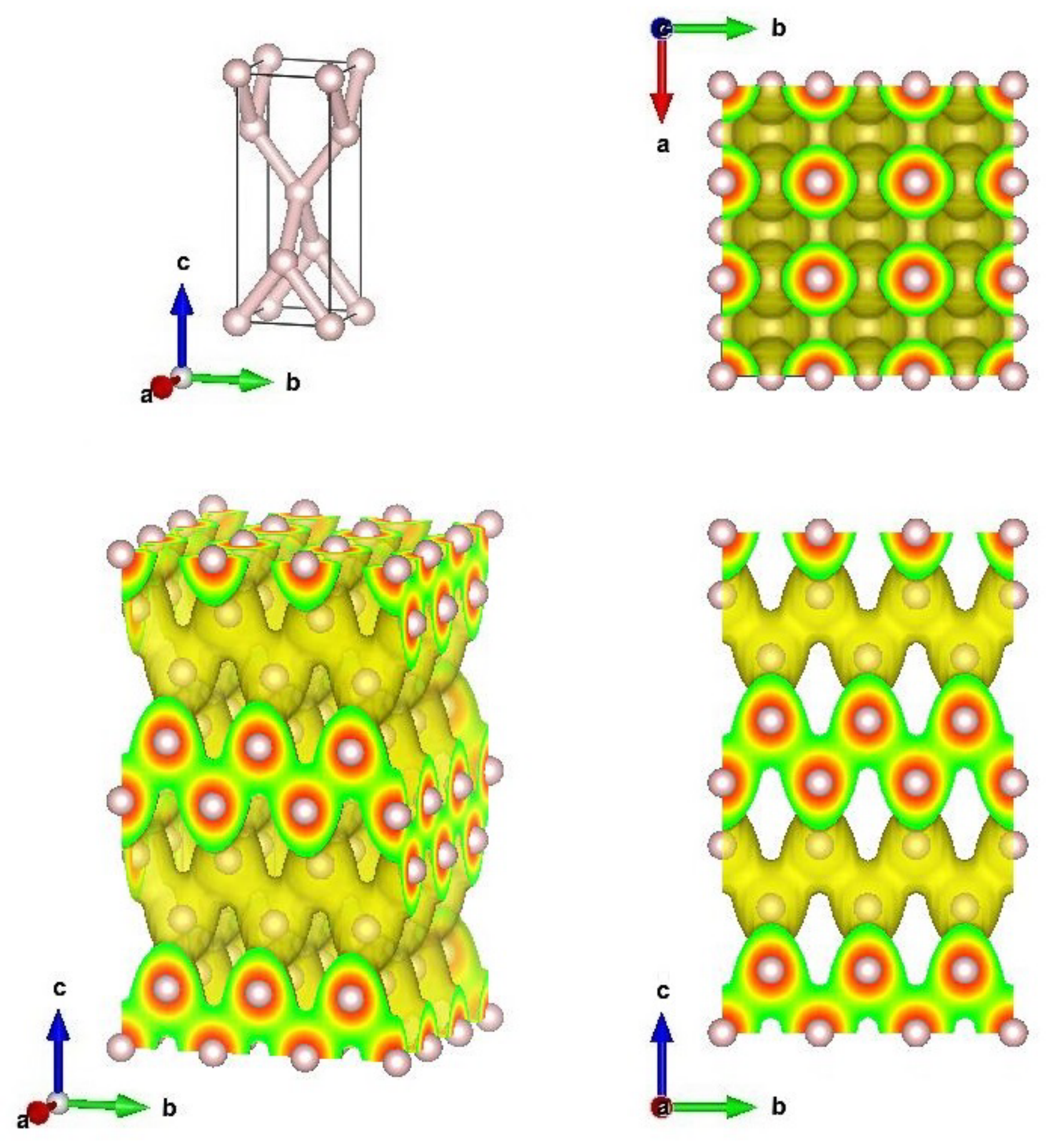}
\caption{Conventional cell of $I4_1/amd$ hydrogen at 495 GPa, and 3D electron localization function (isosurface=0.5) of its 3$\times$3$\times$2 supercell. At the boundary, increasing electron localization from 0.5 to 1 is plotted with colours from green to red.}
\label{f1} 
\end{figure}

We start by structural optimization using the Vienna \textit{ab-initio} simulation package (VASP) based on density functional theory (DFT) \cite{Kresse1996,Kresse1996a} under the generalized gradient approximation (GGA) expressed by the Perdew-Burke-Ernzerhof (PBE) functional \cite{Perdew1996}. A plane-wave basis set with kinetic energy cutoff of 800 eV is employed with 51$\times$51$\times$21 (41$\times$41$\times$41) \textbf{k}-mesh for conventional (primitive) cell at a pressure of 495 GPa during structural relaxation, until the energy differences are converged within 10$^{-6}$ eV, with a Hellman-Feynman force convergence threshold of 10$^{-4}$ eV/\AA.

\begin{table}[h]
\centering
\caption{Calculated X-ray diffraction patterns for the conventional cell of $I4_1/amd$ hydrogen.}
\begin{tabular}{cccccccc}
\hline
 No. & $hkl$ & 2$\theta$ ($^{\circ}$) & $d$ (\AA) & FWHM ($^{\circ}$) \\
\hline
 1 & 001 & 28.6 & 3.1222 & 0.075 \\
 2 & 002 & 59.2 & 1.5611 & 0.081 \\
 3 & 100 & 79.0 & 1.2114 & 0.091 \\
 4 & 011 & 86.1 & 1.1294 & 0.096 \\
 5 & 003 & 95.6 & 1.0407 & 0.104 \\
 6 & 110 & 107.6 & 0.9571 & 0.118 \\
\hline
\end{tabular}
\label{t1}
\end{table}

The conventional cell of the $I4_1/amd$ phase at 495 GPa is shown in Figure~\ref{f1}. The calculated lattice parameters are $a=b=1.2114$ \AA\ and $c=3.1222$ \AA, respectively, which are consistent with other theoretical studies \cite{McMahon2011,Borinaga2016}. The electron localization function maps of $I4_1/amd$ hydrogen at 495 GPa show a typical metallic behavior: Regions with constant 0.5 (corresponding to the electron-gas like pair probability) are connected one to the other by channels, forming infinite three-dimensional networks \cite{Silvi2000}. In order to confirm the structure of the experimentally reported metallic hydrogen, we simulate the X-ray diffraction patterns of the conventional cell \cite{vesta}. The results are shown in Table~\ref{t1}. 

\subsection{Electronic structure}

\begin{figure}[h]
\centering
\includegraphics[width=\linewidth]{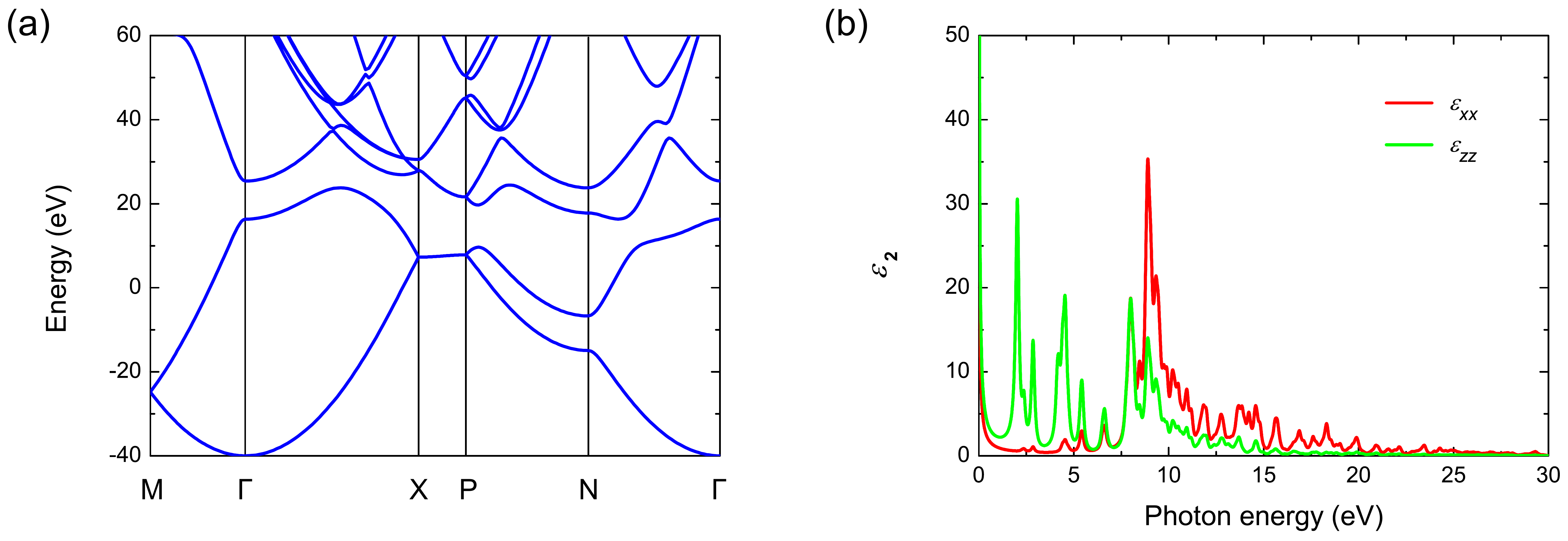}
\caption{(a) Electronic structure and (b) imaginary part of dielectric function for $I4_1/amd$ hydrogen at 495 GPa.}
\label{2}
\end{figure}

Figure~\ref{2}(a) shows the electronic band structure of $I4_1/amd$ hydrogen at 495 GPa. The Fermi level is crossed by five different bands, confirming the metallic behaviour. The huge dispersion is associated to the dominating kinetic term in the energies of the electronic states. Different from the free-electron approximation \cite{Borinaga2016}, the band gap open across the Brillouin zone at about 20 eV due to the interaction of the electrons with the proton lattice.

Borinaga \textit{et al} has pointed out that the electron-electron interactions such as Hartree, exchange and correlation effects give no significant contribution to the band structure \cite{Borinaga2016}. However, after photo-excitation, an electron is excited from valence band into conduction band, leaving behind a positively charged hole. This single-particle excitation cannot be described by non-interacting particles with infinite lifetime. Due to interactions with other particles, the excited electrons and holes become quasiparticles with finite lifetimes. This acquires a self-energy to account for all the electron-electron interactions \cite{Louie2006}. In addition, the negatively charged quasielectron is attracted by the quasihole, stabilizing the electron-hole pair and forming a new quasiparticle, an exciton. Apart from this attraction, a possible electron-hole exchange is also an important many-body effect. Moreover, the electron-hole pairs are weakened by the screening of the electronic system. Therefore, self-energy, excitonic, and other many-electron interaction effects may strongly influence the excited-state properties of metallic hydrogen \cite{Bechstedt2016}.

\subsection{Optical properties}

Subsequent to the standard-DFT results, self-consistent GW$_0$ corrections are undertaken \cite{Shishkin2006,Shishkin2007,Fuchs2007} with eight iterations of $G$. The energy cutoff for the response function is set to be 300 eV. A total of 20 (valence and conduction) bands are used with a \textbf{k}-point sampling of $27\times 27\times 27$. The Bethe-Salpeter equation (BSE) is carried out on top of GW$_0$ calculation with the Tamm-Dancoff approximation \cite{Albrecht1998,Rohlfing1998,Sander2015}. No profound effects on the optical properties are found by including the excitonic effects. Thus we present the GW$_0$ results hereafter.

\begin{table*}[h]
\centering
\caption{Calculated plasma frequency from intraband transitions $\omega_{p}^{intra}$ and interband transitions $\omega_{p}^{inter}$ and electrical conductivity $\sigma$ of $I4_1/amd$ hydrogen at 495 GPa.}
\begin{tabular}{cccc}
\hline
 direction & $\omega_{p}^{intra}$ (eV) & $\omega_{p}^{inter}$ (eV) & $\sigma$ (kS/m) \\
\hline
 $x$ & 23.17 & 24.06 & 1097 \\
 $z$ & 54.61 & 17.20 & 6094 \\
\hline
\end{tabular}
\label{plasma}
\end{table*}

The calculated imaginary part of dielectric function of $I4_1/amd$ hydrogen at 495 GPa is present in Figure~\ref{2}(b). Here we use current-current correlation function instead of density-density correlation function to account for the contribution of Drude terms in metals \cite{Sangalli2012}. The dielectric function diverges at zero frequency because of the free-electron contribution. We also list the plasma frequency from intraband and interband transitions, as well as the electrical conductivity $\sigma$ in Table~\ref{plasma}. The calculated total plasma frequency along the $x$ direction is 33.40 eV, agreeing well with the experimental value of 32.5$\pm$2.1 eV \cite{Dias2017}.

\section{Mechanical properties}

\subsection{Mechanical stability}

To evaluate the mechanical stability of $I4_1/amd$ hydrogen at 495 GPa, we calculate the elastic tensor coefficients of the primitive cell including ionic relaxations using the finite differences method \cite{LePage2002,Wu2005} with a $\Gamma$-centered 34$\times$34$\times$34 \textbf{k}-mesh. The elastic coefficients with the contributions for distortions with rigid ions, the contributions from the ionic relaxations and including both, are present in Table~\ref{t2}.

\begin{table*}[h]
\centering
\caption{Calculated elastic coefficients $C_{ij}$ for the primitive cell of $I4_1/amd$ hydrogen.}
\begin{tabular}{cccccccc}
\hline
 & $C_{11}$ (GPa) & $C_{33}$ (GPa) & $C_{44}$ (GPa) & $C_{66}$ (GPa) & $C_{12}$ (GPa) & $C_{13}$ (GPa) \\
\hline
 Rigid ions & 1974.22 & 1249.60 & 111.33 & 18.54 & -317.16 & 598.73 \\
 Ionic relaxations & -13.74 & 0.0 & -45.57 & 0.0 & 13.74 & 0.0 \\
 Total & 1960.48 & 1249.60 & 65.76 & 18.54 & -303.42 & 598.73 \\
\hline
\end{tabular}
\label{t2}
\end{table*}

$I4_1/amd$ hydrogen is tetragonal crystal system with $a=b\neq c$. According to Born-Huang's lattice dynamical theory \cite{Born1954,Wu2007}, the mechanical stability conditions for tetragonal phase are given by $C_{11} > 0,\ C_{33} > 0,\ C_{44} > 0,\ C_{66} > 0,\ (C_{11} - C_{12}) > 0,\ (C_{11} + C_{33} - 2C_{12}) > 0,\ 2(C_{11} + C_{12}) + C_{33} + 4C_{13} > 0$. The calculated elastic constants of metallic hydrogen satisfy the corresponding Born stability criteria, indicating the $I4_1/amd$ phase is mechanically stable.

\subsection{Bulk modulus and shear modulus}

The bulk modulus $B$ describes the material's response to uniform hydrostatic pressure. For tetragonal structure, the Voigt and Reuss methods are used to evaluate $B$ \cite{Wu2007,Bannikov2012,Yu2015a}
\begin{equation}
B_V = \frac{2(C_{11} + C_{12}) + C_{33} + 4C_{13}}{9},
\end{equation}
\begin{equation}
B_R = \frac{(C_{11} + C_{12})  C_{33} - 2C_{13}^2}{C_{11} + C_{12} + 2C_{33} - 4C_{13}}.
\end{equation}
The shear modulus $G$ the material's response to shear stress, and can be given by
\begin{equation}
G_V = \frac{4C_{11} - 2C_{12} + 2C_{33} - 4C_{13} + 12C_{44} + 6C_{66}}{30},
\end{equation}
\begin{equation}
G_R = \frac{15}{18B_V/[(C_{11} + C_{12})  C_{33} - 2C_{13}^2] + 6/(C_{11}-C_{12}) + 6/C_{44} + 3/C_{66}}.
\end{equation}
The Hill method calculations are as follows:
\begin{equation}
B_H = \frac{1}{2} (B_V + B_R),
\end{equation}
\begin{equation}
G_H = \frac{1}{2} (G_V + G_R).
\end{equation}
The calculated bulk modulus and shear modulus are shown in Table~\ref{t3}.

\begin{table*}[h]
\centering
\caption{Calculated bulk modulus $B$, shear modulus $G$, $B/G$ ratio, Young's modulus $E$, Poisson's ratio, and anisotropy index $A$ for the primitive cell of $I4_1/amd$ hydrogen.}
\begin{tabular}{cccccccccccc}
\hline
 $B_V$ (GPa) & $B_R$ (GPa) & $B_H$ (GPa) & $G_V$ (GPa) & $G_R$ (GPa) & $G_H$ (GPa) & $B_H/G_H$ & $E$ (GPa) & $\nu$ & $A^U$ & $A^B$ & $A^G$ \\
\hline
 773.18 & 768.56 & 770.87 & 315.11 & 56.39 & 185.75 & 4.15 & 515.83 & 0.39 & 22.94 & 0.30\% & 69.64\% \\
\hline
\end{tabular}
\label{t3}
\end{table*}

The bulk modulus of metallic hydrogen is even higher than diamond (443 GPa) and cubic C$_3$N$_4$ (496 GPa). This result can be easily explained by the extreme high synthesis pressure. The shear modulus, which represents the resistance to shear deformation against external forces, is much lower than the bulk modulus. This implies that the shear modulus limits the mechanical stability of hydrogen.

The $B/G$ ratio measures the malleability of materials. A high value represents ductility, while a low value is associated with brittleness. The critical value separating ductile and brittle materials is approximately 1.75 \cite{Yu2015a}. Our calculated $B/G$ ratio is 4.15, revealing that metallic hydrogen is ductile.

\subsection{Young's modulus and Poisson's ratio}

The Young's modulus can be evatulated from $B_H$ and $G_H$ \cite{Bannikov2012}
\begin{equation}
E = \frac{9B_H G_H}{3 B_H + G_H},
\end{equation}
and the Poisson's ratio is
\begin{equation}
\nu = \frac{3B_H - 2G_H}{2 (3 B_H + G_H)}.
\end{equation}
The calculated Young's modulus and Poisson's ratios are present in Table~\ref{t3}.

Young's modulus is a mechanical property of linear elastic solid materials, and measures the stiffness of a solid material. A material with a higher Young's modulus is stiffer, which needs more force to deform compared to a soft material. The high Young's modulus indicates that metallic hydrogen is a stiff material.

The Poisson's ratio describes the response in the direction orthogonal to uniaxial strain. Poisson's ratio close to 0 indicates very little lateral expansion when compressed, while a Poisson's ratio of exactly 0.5 represents a perfectly incompressible material deformed elastically at small strains. For $I4_1/amd$ hydrogen, a Poisson's ratio of 0.39 suggests it is less compressible.

\subsection{Elastic anisotropy}

In most single crystals, the elastic response is usually anisotropic. Elastic anisotropy exhibits a different bonding character in different directions and is important in diverse applications such as phase transformations, dislocation dynamics. The universal anisotropy index, which represents a universal measure to quantify the single crystal elastic anisotropy, is defined as \cite{Bannikov2012,Ranganathan2008}
\begin{equation}
A^U = 5\frac{G_V}{G_R} + \frac{B_V}{B_R} - 6,
\end{equation}
$A^U$ is identically zero for locally isotropic single crystals. The departure of $A^U$ from zero defines the extent of single crystal anisotropy. Another way implies the estimation in compressibility and shear
\begin{equation}
A^B = \frac{B_V-B_R}{B_V+B_R},
\end{equation}
\begin{equation}
A^G = \frac{G_V-G_R}{G_V+G_R}.
\end{equation}
A value of zero denotes elastic isotropy and a value of 100\% represents the largest anisotropy. The calculated anisotropy indexes are obtained in Table~\ref{t3}. 

According to universal anisotropy index $A^U$, the average elastic anisotropy is more than two times higher than Li \cite{Ranganathan2008}. Comparing $A^B$ and $A^G$, there is much more anisotropy in shear than in compressibility.

\subsection{Sound velocity and Debye temperature}

\begin{table*}[h]
\centering
\caption{Calculated longitudinal sound velocity $v_l$, transverse sound velocity $v_t$ for the primitive cell of $I4_1/amd$ hydrogen for [100], [001] and [110] directions.}
\begin{tabular}{cccccccccccc}
\hline
 [100] & (km/s) & & [001] & (km/s) & & [110] & (km/s) \\
 $[100]v_l$ & $[001]v_{t1}$ & $[010]v_{t2}$ & $[001]v_l$ & $[100]v_{t1}$ & $[010]v_{t2}$ & $[110]v_l$ & $[001]v_{t1}$ & $[1\bar{1}0]v_{t2}$ \\
\hline
 36.65 & 6.71 & 3.56 & 29.26 & 3.56 & 3.56 & 24.09 & 6.71 & 27.85 \\
\hline
\end{tabular}
\label{t4}
\end{table*}

The sound velocity is determined by the symmetry of the crystal and the propagation direction. The tetragonal symmetry dictates that pure transverse and longitudinal modes can only exist for [100], [001] and [110] directions. In all other directions the propagating waves are either quasi-transverse or quasi-longitudinal. In the principal directions the acoustic velocities can be simply written as \cite{Feng2012}

\noindent For [100] direction:
\begin{equation}
[100]v_l = \sqrt{C_{11}/\rho},\ \ [001] v_{t1} = \sqrt{C_{44}/\rho},\ \ [010] v_{t2} = \sqrt{C_{66}/\rho};
\end{equation}

\noindent For [001] direction:
\begin{equation}
[001]v_l = \sqrt{C_{33}/\rho},\ \ [100] v_{t1} = [010] v_{t2} = \sqrt{C_{66}/\rho};
\end{equation}

\noindent For [110] direction:
\begin{equation}
[110]v_l = \sqrt{(C_{11}+C_{12}+2C_{66})/\rho},\ \ [001] v_{t1} = \sqrt{C_{44}/\rho},\ \ [1\bar{1}0] v_{t2} = \sqrt{(C_{11}-C_{12})/2\rho}.
\end{equation}
The calculated sound velocities are presented in Table~\ref{t4}. 

\begin{table}[h]
\centering
\caption{Calculated average longitudinal sound velocity $v_l$, transverse sound velocity $v_t$, average sound velocity $v_s$ and Debye temperature $\theta_D$ for the primitive cell of $I4_1/amd$ hydrogen.}
\begin{tabular}{ccccccccc}
\hline
 $v_l$ (km/s) & $v_t$ (km/s) & $v_s$ (km/s) & $\theta_D$ (K) \\
\hline
 26.42 & 11.28 & 12.75 & 3626.64 \\
\hline
\end{tabular}
\label{t5}
\end{table}

The Debye temperature $\theta_D$ can be calculated as follows \cite{Shao2016}
\begin{equation}
\label{Debye-stiffness}
\theta_D=\frac{h}{k_B} \bigg( \frac{3N}{4\pi V} \bigg)^{1/3} v_s,
\end{equation}
where $h$ is the Planck constant, $k_B$ is the Boltzmann constant, $N$ is the number of atoms in the cell, $V$ is the volume of the unit cell, and $v_s$ is the average sound velocity given by 
\begin{equation}
v_s=\bigg[\frac{1}{3} \big( \frac{1}{v_l^3} + \frac{2}{v_t^3} \big) \bigg]^{-1/3},
\end{equation}
The average sound velocity in crystals can be determined from $B$ and $G$ \cite{Yu2015a},
\begin{equation}
v_l=\sqrt{\frac{3B+4G}{3\rho}},
\end{equation}
\begin{equation}
v_t=\sqrt{\frac{G}{\rho}}.
\end{equation}
The calculated sound velocities and Debye temperature are shown in Table~\ref{t5}.

The speed of sound can be used to measure the speed of phonons propagating through the lattice. The ultrahigh longitudinal sound velocity of metallic hydrogen indicates high phonon velocity. The Debye temperature measures the temperature above which all modes begin to be excited \cite{Nakashima1992}. Thus a high $\theta_D$ indicates weak three-phonon scattering and hence a high lattice thermal conductivity.

\section{Phonon properties}

\subsection{Dynamical stability}
The Born-Huang mechanical stability criteria provide a necessary condition for the dynamical stability, but not a sufficient one \cite{Zhou2014a}. Therefore we need to examine the dynamical stability of $I4_1/amd$ hydrogen at 495 GPa by calculating the phonon dispersion of the primitive cell. 

The harmonic interatomic force constants are obtained using density functional perturbation theory (DFPT) within a supercell approach \cite{DFPT}. A 7$\times$7$\times$7 supercell with 5$\times$5$\times$5 \textbf{q}-mesh is used. The phonon dispersion and thermodynamical properties are calculated from the interatomic force constants using the PHONOPY code \cite{Togo2008,Togo2015}.

\begin{figure*}[h]
\centering
\includegraphics[width=\linewidth]{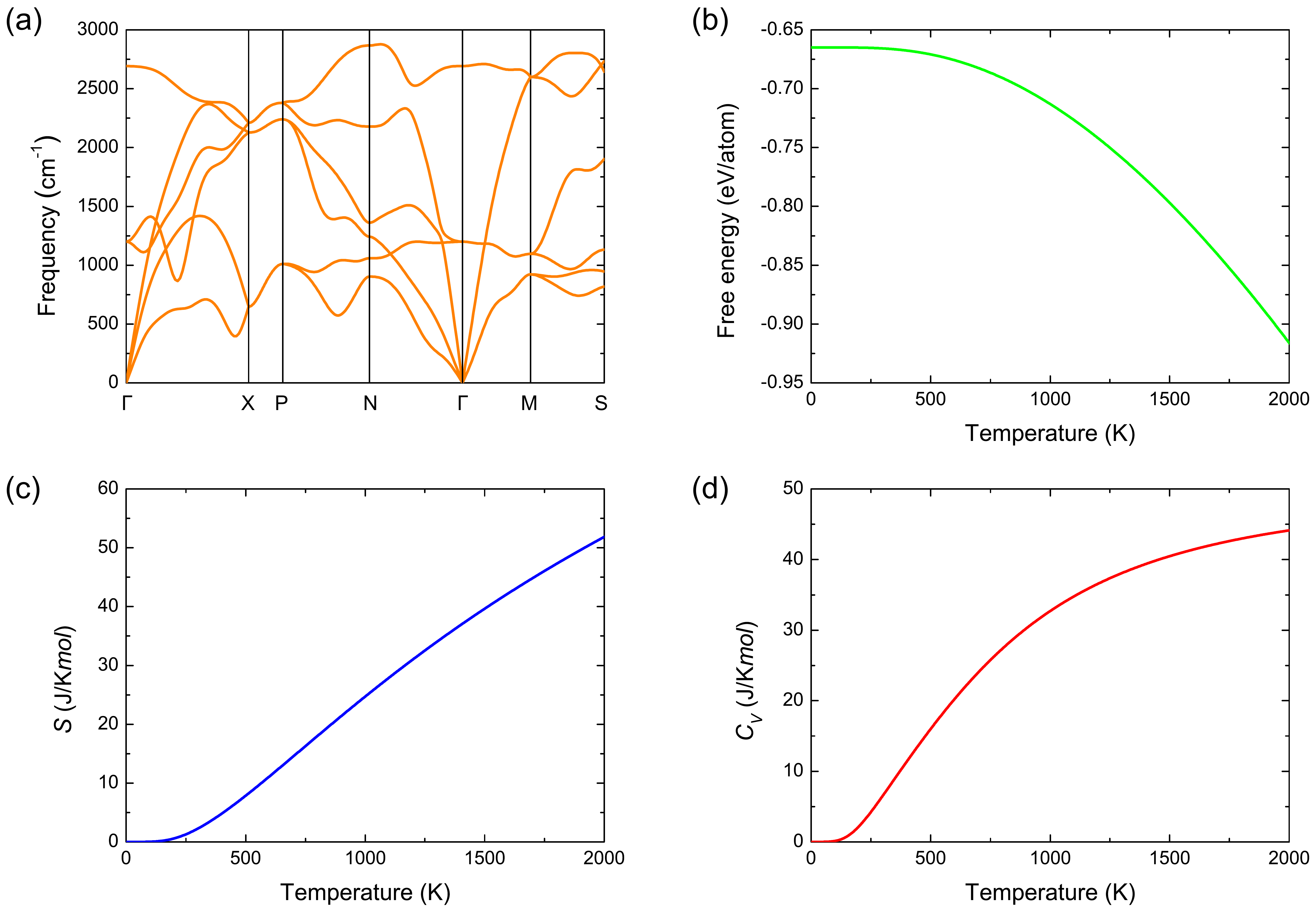}
\caption{(a) Phonon dispersion of $I4_1/amd$ hydrogen at 495 GPa. (b) Helmholtz free energy, (c) entropy, and (d) heat capacity for $I4_1/amd$ hydrogen at 495 GPa as a function of temperature.}
\label{f2} 
\end{figure*}

As shown in Figure~\ref{f2}(a), no imaginary frequencies exist in the whole Brillouin zone, indicating dynamical stability at 0 K. The calculated Raman active modes are 1200.24 cm$^{-1}$ ($E_g$) and 2692.70 cm$^{-1}$ ($B_{1g}$) respectively, which can be used as a reference for characterization of $I4_1/amd$ hydrogen at 495 GPa. The calculated phonon dispersion agrees well with previous result \cite{McMahon2011}.

\subsection{Thermodynamical properties}

For light elements such as hydrogen, phonons play an important role in determining the thermodynamical properties of crystals both at 0 K and at finite temperatures \cite{Setten2007}. Using phonon frequencies in the Brillouin zone, we further examine the thermodynamical properties of metallic hydrogen by calculating the Helmholtz free energy $F$ \cite{Setten2007},
\begin{equation}
F = E_{tot} + \frac{1}{2} \sum\limits_{\textbf{q}j}\hbar\omega_{\textbf{q}j}+k_BT\sum\limits_{\textbf{q}j}\ln[1-\exp(-\hbar \omega_{\textbf{q}j}/k_BT)],
\end{equation}
where $E_{tot}$ is the total energy of the crystal, $\hbar$ is the reduced Planck constant, $\omega_{\textbf{q}j}$ is the phonon frequency of the $j$-th branch with wave vector \textbf{q}, $T$ is the temperature, and the summation term is the Helmholtz free energy for phonons \cite{Togo2008,Togo2015}. The first summation term is a temperature-free term corresponding to the zero point energy of phonons; and the second summation term is a temperature-dependent term referring to the thermally induced occupation of the phonon modes. The calculated zero point energy of $I4_1/amd$ hydrogen at 495 GPa is 0.295 eV/atom, which is the Helmholtz free energies of phonons at 0 K. Temperature is also an important thermodynamic variable for determining the stability of materials. The Helmholtz free energies $F$ as a function of temperature are shown in Figure~\ref{f2}(b). At higher temperature, the phonon modes are occupied according to Bose-Einstein statistics, and the free energy further decreases.

From the Helmholtz free energy, the other thermodynamical behavior can be deduced \cite{Born1954}. The entropy is
\begin{equation}
S=-\frac{\partial F}{\partial T}.
\end{equation}
The calculated entropies for the four structures as a function of temperature are shown in Fig.~\ref{f2}(c). The difference between the entropies of different phases can be used to determine the relative stability \cite{Pavone1998,McMahon2011a}.

The isometric heat capacity can be calculated as
\begin{equation}
{C_V}=\sum_{\textbf{q}j}k_B\left(\frac{\hbar\omega_{\textbf{q}j}}{k_BT}\right)^2\frac{\exp(\hbar\omega_{\textbf{q}j}/k_BT)}{[\exp(\hbar\omega_{\textbf{q}j}/k_BT)-1]^2}.
\end{equation}
The calculated heat capacities per atom are shown in Figure~\ref{f2}(d), which approach their Dulong-Petit classical limit (2$\times$24.94 J/K$mol$) at high temperatures.

\subsection{Lattice anharmonicity}

Accurate simulation of anharmonic properties such as thermomechanics and thermal expansion is important for understanding thermal behaviours in solids and their realistic applications. The anharmonic properties can be derived from the volume dependences of phonon dispersion using the quasiharmonic approximation method \cite{Togo2008,Togo2015}, and temperature is assumed to indirectly affect vibrational properties via thermal expansion. Using the same supercell and \textbf{q}-mesh in previous phonon calculations, we calculate the phonon spectra of ten volumes, and the thermal expansion and thermomechanics are derived by fitting the free energy-volume relationship.

\begin{figure*}[h]
\centering
\includegraphics[width=\linewidth]{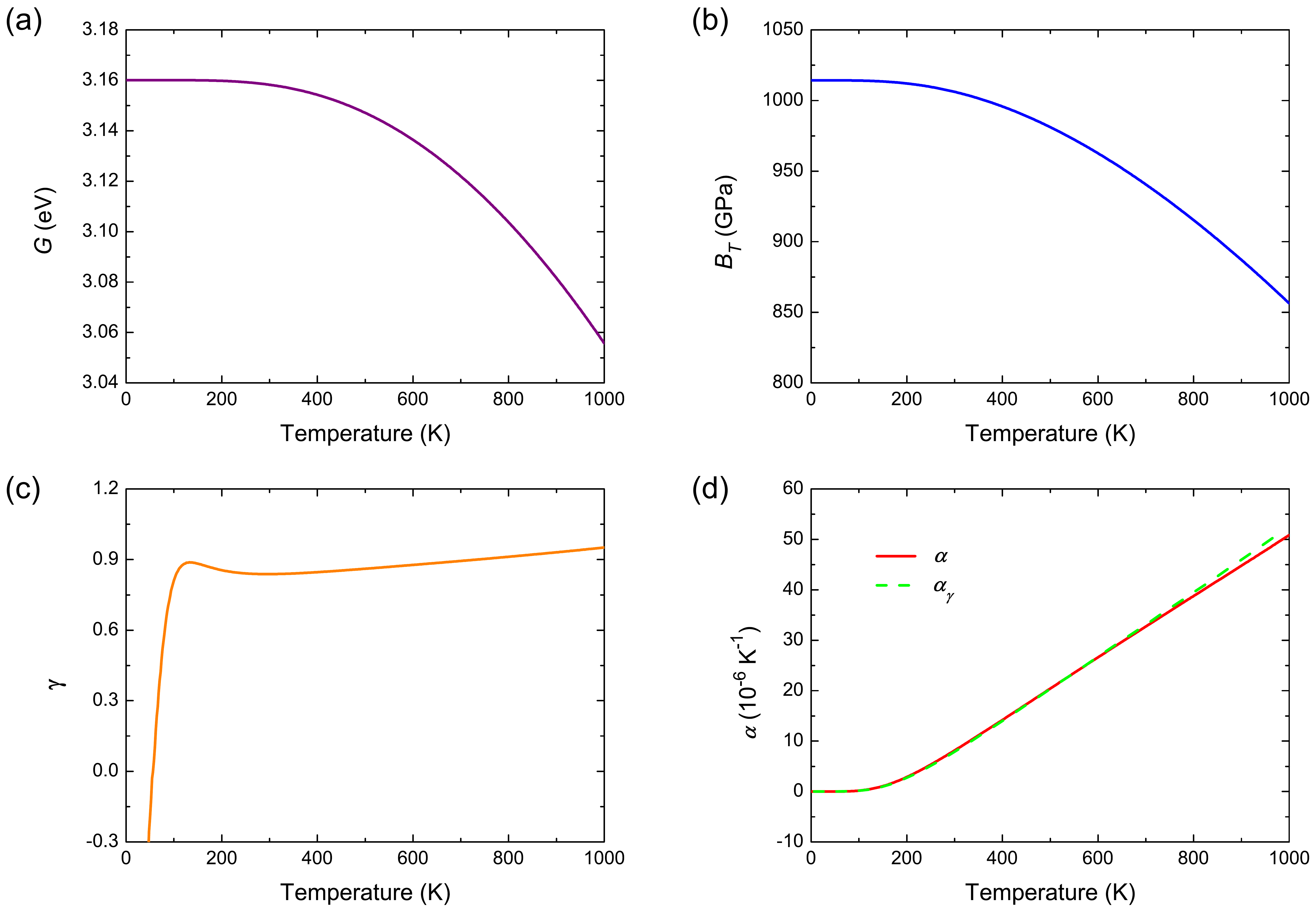}
\caption{(a) Gibbs free energy, (b) isothermal bulk modulus, (c) Gr\"uneisen parameter, and (d) volume thermal expansion coefficient of $I4_1/amd$ hydrogen at 495 GPa as a function of temperature.}
\label{f3} 
\end{figure*}

Gibbs free energy $G$ at given temperature $T$ and pressure $p$ is obtained from Helmholtz free energy $F$ via finding a minimum value by changing volume $V$ \cite{Togo2015},
\begin{equation}
G(T,p)=\textrm{Min}_V[F(T;V)+pV],
\end{equation}
where $F$ is the sum of electronic internal energy and phonon Helmholtz free energy. The calculated $G$ is depicted in Figure~\ref{f3}(a). The minimum value is obtained by the fits to third-order Birch-Murnaghan equation of states.

The calculated isothermal bulk modulus is present in Figure~\ref{f3}(b). An anticipated thermal softening is observed in the temperature dependence of $B_T$. From 0 to 1000 K, the softening of the isothermal bulk modulus for metallic hydrogen is 157.98 GPa, which is much larger than those in C and Al$_2$O$_3$ \cite{Huang2016}.

The Gr\"uneisen parameter $\gamma$ describes the effect of thermal expansion on vibrational properties, and provides information on the anharmonic interactions. A larger Gr\"uuneisen parameter indicates stronger anharmonic vibrations. As shown in Figure~\ref{f3}(c), the calculated $\gamma$ at room temperature is 0.84.

The volume thermal expansion coefficient $\beta$ can be calculated from the Gr\"uneisen parameter \cite{Grimvall}
\begin{equation}
\beta = \frac{\gamma C_V}{B_T V}.
\end{equation}
Another approach is to obtain $\beta$ from the calculated equilibrium volumes $V$ at temperature $T$
\begin{equation}
\beta = \frac{1}{V(T)}\frac{\partial V(T)}{\partial T}.
\end{equation}
As shown in Figure~\ref{f3}(d), the $\beta$ curves from these methods are in excellent agreement with each other. It should be noticed that it is challenging to measure $\beta$ accurately in experiment because thermal expansion varies with the crystalline orientation and there may be some metastable phases of metallic hydrogen at relative high temperatures.

\subsection{Phonon transport}

The lattice thermal conductivity $\kappa$ can be calculated using the self-consistent iterative approach \cite{Omini1996,ShengBTE,Li2012a,Li2012,Li2013a} as a sum of contribution of all the phonon modes $\lambda$,
\begin{equation}\label{kappa-eq}
\kappa^{\alpha\alpha}=\frac{1}{V}\sum_{\textbf{q}j}C_{\textbf{q}j} \tau_{\textbf{q}j} {v_{\textbf{q}j}^{\alpha}}^2,
\end{equation}
where $C_{\textbf{q}j}$ is the heat capacity per mode, $\tau_{\textbf{q}j}$ is the mode relaxation time, and $v_{\textbf{q}j}^{\alpha}$ is the group velocity along $\alpha$ direction, respectively. The lattice dynamical properties $C_{\textbf{q}j}$ and $v_{\textbf{q}j}^{\alpha}$ in Eq.~(\ref{kappa-eq}) can be obtained by the phonon dispersion relation with harmonic interatomic force constants as input, while the $\tau_{\textbf{q}j}$ provides information about the anharmonic interactions and can be obtained using the anharmonic interatomic force constants. The anharmonic interatomic force constants are calculated using a supercell-based, finite-difference method \cite{Li2012}, and a 7$\times$7$\times$7 supercell is used with a 5$\times$5$\times$5 \textbf{q}-mesh. We include the interactions with the 40th nearest-neighbour atoms. A discretizationa of the Brillouin zone into a $\Gamma$-centered regular grid of 32$\times$32$\times$32 $\textbf{q}$ points is introduced with scale parameter for broadening chosen as 0.1. When the phonon $\textbf{q}$-points mesh increases from 28$\times$28$\times$28 to 32$\times$32$\times$32, the difference of $\kappa$ between the two grids is less than 1\%, verifying the convergence of our calculations.

\begin{figure*}[h]
\centering
\includegraphics[width=\linewidth]{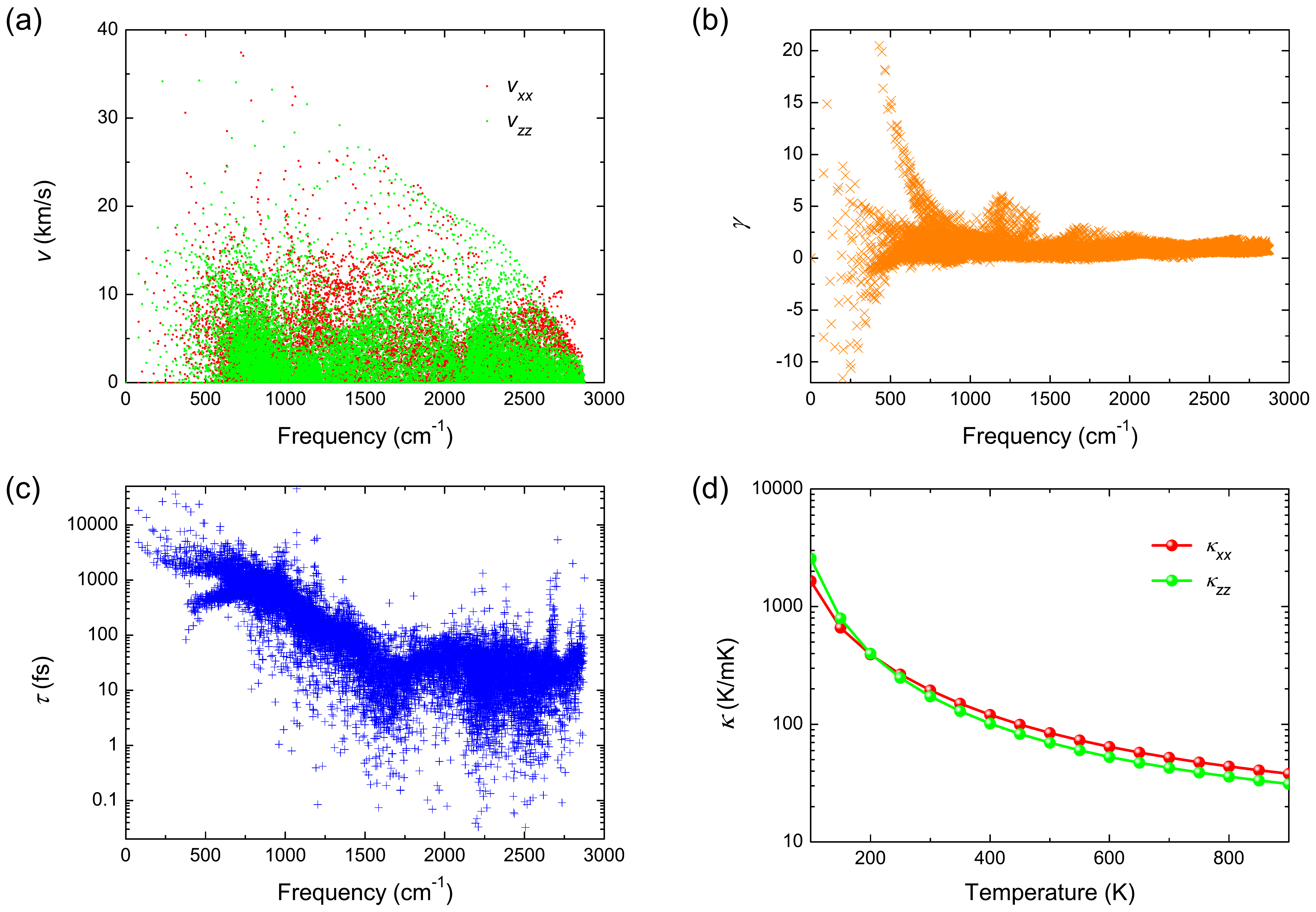}
\caption{(a) Phonon group velocity, (b) mode Gr\"uneisen parameters, (c) relaxation time of $I4_1/amd$ hydrogen at 495 GPa at 300 K, and (d) lattice thermal conductivity of $I4_1/amd$ hydrogen at 495 GPa as a function of temperature.}
\label{f4} 
\end{figure*}

Figure~\ref{f4}(a) presents the phonon group velocities calculated from the phonon spectrum within the whole Brillouin zone. The group velocity shows strong anisotropy along the $x$ and $z$ directions. The phonon group velocities in long-wavelength limit for the acoustic modes agree well with the sound velocities calculated from elastic coefficients.

The phonon scattering depends on two factors: the number of channels available for a phonon to get scattered, and the strength of each scattering channel. The former factor is determined by whether there exist three phonon groups that can satisfy both energy and quasi-momentum conservations. The latter factor depends on the anharmonicity of a phonon mode, which is described by the mode Gr\"uneisen parameter. The mode Gr\"uneisen parameters provide information about anharmonic interactions, and can be calculated directly from the anharmonic interatomic force constants. Figure~\ref{f4}(b) shows the mode Gr\"uneisen parameters. Strong anisotropy is found in the Gr\"uneisen parameters at low frequencies. To provide more physical insight, we investigate the phonon relaxation time of each phonon mode at 300 K in Figure~\ref{f4}(c). The relaxation times are highly anisotropic at low frequencies as well.

The thermal conductivity for $I4_1/amd$ hydrogen at 495 GPa at 300 K are higher than 170 W/mK along the $x$ and $z$ directions, as listed in Table~\ref{thermal}. We extract the contributions of different phonon branches to $\kappa$ along the $x$ and $z$ directions, as shown in Table~\ref{thermal}. Here the different modes are simply distinguished by frequency \cite{Peng2016}. For phonon transport along $x$ direction, the contribution of TA$_2$ phonons is largest among all phonon modes. For phonon transport along $z$ direction, TA$_1$ phonons contribute the most to $\kappa$. For both directions, the acoustic phonons dominate the heat transport.

\begin{table*}[h]
\centering
\caption{$\kappa$ at 300 K and contribution of different phonon modes branches (TA$_1$, TA$_2$, LA, and all optical phonons) towards the $\kappa$ of $I4_1/amd$ hydrogen at 495 GPa.}
\begin{tabular}{cccccc}
\hline
 direction & $\kappa$ (W/mK) & TA$_1$ (\%) & TA$_2$ (\%) & LA (\%) & Optical (\%) \\
\hline
 $x$ & 194.72 & 28.6 & 46.8 & 24.3 & 0.3 \\
 $z$ & 172.96 & 55.6 & 22.7 & 21.4 & 0.3 \\
\hline
\end{tabular}
\label{thermal}
\end{table*}

The intrinsic lattice thermal conductivity $\kappa$ of $I4_1/amd$ hydrogen at 495 GPa is present in Figure~\ref{f4}(d). The $\kappa$ is weakly anisotropic ($\kappa_{xx}=\kappa_{yy}\neq\kappa_{zz}$). The anisotropy in thermal transport can be attributed to the anisotropic natures of phonon dispersion of acoustic modes, which results in the orientation-dependent group velocities and Gr\"uneisen parameters as mentioned above. The orientation-dependence of thermal transport properties provides a way to determine the optimized transport directions for potential applications.

\section{Conclusion}

State-of-the-art computational material techniques enable a systematic investigation of different properties of metallic hydrogen. The calculated electronic structure and dielectric function confirm the metallic behaviour of $I4_1/amd$ hydrogen at 495 GPa. The calculated total plasma frequency from intraband and interband transitions of 33.40 eV agrees well with the experimental value. The phase is mechanical stable, and the stability is limited by shear modulus other than bulk modulus. The high Young's modulus indicates that $I4_1/amd$ hydrogen is a stiff material, while the Poisson's ratio suggests it is less compressible. The elastic anisotropy, sound velocity and Debye temperature are also investigated in detail. The dynamical stability of $I4_1/amd$ hydrogen is confirmed by phonon dispersion. The thermodynamical properties, as well as lattice anharmonicity are also calculated to understand thermal behaviours in metallic hydrogen. The lattice thermal conductivity of $I4_1/amd$ hydrogen is 194.72 W/mK and 172.96 W/mK along the $x$ and $z$ directions, respectively, where acoustic phonons dominate heat transport. Our results show the power of first-principles calculations in predicting a variety of material properties in extreme conditions where purely experimental approaches are impractical, and pave way for potential applications of metallic hydrogen at extremely high pressure and high temperature.


\section*{Acknowledgement}

This work is supported by the National Natural Science Foundation of China under Grants No. 11374063 and 11404348, and the National Basic Research Program of China (973 Program) under Grant No. 2013CBA01505.

\section*{References}
\bibliography{new}

\end{document}